\documentclass[lettersize,journal]{IEEEtran}
\usepackage{amsmath,amsfonts,amssymb}
\usepackage{algorithmic}
\usepackage{algorithm}
\usepackage{array}
\usepackage{textcomp}
\usepackage{stfloats}
\usepackage{url}
\usepackage{multirow}
\usepackage{tabularx,booktabs} 
\usepackage{verbatim}
\usepackage{graphicx}
\usepackage{cite}
\usepackage{makecell}
\usepackage{url}
\usepackage{relsize} 
\usepackage{color}
\usepackage{graphicx}
\usepackage{float}
\usepackage{subfigure}
\usepackage{enumitem}
\usepackage{bbm}
\usepackage{amsmath}
\usepackage{stfloats}
\makeatletter
\renewcommand{\maketag@@@}[1]{\hbox{\m@th\normalsize\normalfont#1}}%
\makeatother
\usepackage{hyperref}  

\begin{document}

\title{UAV-Enabled Wireless-Powered Underground Communication Networks: A Novel Time Allocation Approach}

\author{Kaiqiang Lin,~\IEEEmembership{Member,~IEEE,}
Yijie Mao,~\IEEEmembership{Member,~IEEE,}
Onel Luis Alcaraz López,~\IEEEmembership{Senior Member,~IEEE,} and
Mohamed-Slim Alouini,~\IEEEmembership{Fellow,~IEEE}

\thanks{This work was supported in part by the National Nature Science Foundation of China under Grant 62201347, Shanghai Sailing Program under Grant 22YF1428400, and Research Council of Finland (Grants 348515 (UPRISING) and 369116 (6G Flagship)). (\textit{Corresponding author: Yijie Mao})}  

\thanks{K. Lin and M.-S. Alouini are with the Division of Computer, Electrical and Mathematical Sciences and Engineering, King Abdullah University of Science and Technology, Saudi Arabia (E-mail: kaiqiang.lin@kaust.edu.sa; slim.alouini@kaust.edu.sa).}

\thanks{Y. Mao is with the School of Information Science and Technology, ShanghaiTech University, Shanghai, China (E-mail: maoyj@shanghaitech.edu.cn).}

\thanks{O. L. A. López is with the Centre for Wireless Communications, University of Oulu, Finland (Email: onel.alcarazlopez@oulu.fi).}
}
\maketitle
\begin{abstract}
Wireless-powered underground communication networks (WPUCNs), which allow underground devices (UDs) to harvest energy from wireless signals for battery-free communication, offer a promising solution for sustainable underground monitoring. However, the severe wireless signal attenuation in challenging underground environments and the costly acquisition of channel state information (CSI) make large-scale WPUCNs economically infeasible in practice. To address this challenge, we introduce flexible unmanned aerial vehicles (UAVs) into WPUCNs, leading to UAV-enabled WPUCN systems. In this system, a UAV is first charged by a terrestrial hybrid access point (HAP), then flies to the monitoring area to wirelessly charge UDs. Afterwards, the UAV collects data from the UDs and finally returns to the HAP for data offloading. Based on the proposed UAV-enabled WPUCN system, we first propose its energy consumption model and a hybrid wireless energy transfer (WET) approach (i.e., UDs can harvest energy from both the HAP and the UAV) relying on full-CSI and CSI-free multi-antenna beamforming. Then, we formulate and address a time allocation problem to minimize the energy consumption of UAV, while ensuring that the throughput requirements of all UDs are met and all sensor data is offloaded. Through simulations of a realistic farming scenario, we demonstrate that the proposed hybrid WET approach outperforms other WET approaches, with performance gains influenced by the number of antennas, communication distance, number of UDs, and underground conditions. Additionally, under the optimized time allocation, we found that the proposed hybrid WET approach based on a CSI-free multi-antenna scheme achieves the lowest UAV's energy consumption among all WET mechanisms, thereby enabling sustainable underground monitoring in WPUCNs.
\end{abstract}


\section{Introduction}
\IEEEPARstart{W}{ireless} underground sensor networks (WUSNs) enable \textit{in-situ} and real-time monitoring of various underground entities through wirelessly connected underground devices (UDs), facilitating a range of applications, including smart agriculture~\cite{LinMag}, underground infrastructure monitoring~\cite{LinAdhoc}, post-disaster rescue operations~\cite{LiuWETMag}, and border patrol~\cite{WUSNReview}. However, compared to terrestrial wireless sensor networks, UDs with limited battery capacity require more energy to ensure reliable communication in the harsh underground environments~\cite{VuranWUSNsreview}. Moreover, it is often impractical to regularly replace the batteries of UDs. 

To establish sustainable underground monitoring, the radio-frequency (RF) wireless energy transfer (WET) technology has been introduced in WUSNs, giving rise to wireless-powered underground communication networks (WPUCNs)~\cite{LiuWPUSNs}. Several studies have explored techniques for enabling and enhancing WPUCNs~\cite{LiuWPUSNs, LiuMIMOWPUSNs, LinBSWPUSNs, LinCSIfree}. Specifically, the pioneering work of Liu et al.~\cite{LiuWPUSNs} conceptualized a multi-user WPUSN, where the UDs harvest energy from an aboveground power source (PS) via WET to enable wireless information transfer (WIT) to the access point. Herein, the time allocation was designed to maximize network throughput while ensuring communication reliability and diverse data traffic demands. To further improve the throughput performance in highly heterogeneous underground environments, recent work~\cite{LiuMIMOWPUSNs} applied multi-antenna techniques to WPUCNs and designed the optimal beamforming based on the estimated channel state information (CSI). In~\cite{LinBSWPUSNs}, the authors incorporated backscatter communication technology into WPUCNs to support urgent data transmission and enhance resource utilization. Therein, the aim was to allocate time for WET and backscattering to maximize the network throughput with communication reliability assurance. However, these studies assume the availability of accurate CSI, whose acquisition is  inherently imperfect, affected by interference and access collisions, incurs increased effective delay in multi-user networks, and excessive energy consumption for low-power deployments, thereby diminishing the potential benefits of CSI exploitation in WPUCNs. To address these challenges, several promising CSI-free multi-antenna WET schemes, hereinafter referred to just as CSI-free schemes, were considered in~\cite{LinCSIfree}, for powering a large number of nearby UDs. The authors demonstrated the feasibility of several CSI-free schemes, including ``switching antennas" (SA)~\cite{CSIfreeIoT}, ``all antennas transmitting independent signals" (AAIS)~\cite{CSIfreeTcom}, ``all antennas transmitting the same signal" (AASS)~\cite{CSIfreeIoT}, and ``rotary antenna beamforming" (RAB)~\cite{RAB}, in massive WPUCN scenarios by considering practical power budgets and two PS deployment strategies.

Despite extensive efforts on improving the performance of WPUCNs, enabling large-scale and sustainable underground monitoring remains challenging due to high attenuation in underground soil, a high probability of non-line-of-sight (NLOS) conditions, and increased air path losses when UDs are located far from the PS. Although numerous PSs can be deployed to ensure reliable WET operation, this is economically infeasible for large-scale WPUCN scenarios. Instead, unmanned aerial vehicles (UAVs) may be a suitable alternative due to their flexibility, decreasing expense and increasing functionality. Indeed, they may be dispatched to charge remote sensors, collect data from these sensors, or perform both tasks simultaneously~\cite{UAVReview2}. Motivated by these advancements, we propose mounting a hybrid communication and power transfer module on UAVs as a cost-effective solution for wirelessly charging the entire underground area and collecting sensor data from all UDs through shorter-distance line-of-sight (LOS) communication channels.

There has been a growing interest in studying the feasibility of UAV-enabled WET and data collection in terrestrial wireless powered communication networks (WPCNs). It has been demonstrated in~\cite{UAVWETTVT, UAVWET1, UAVWET2} that UAV-enabled WET systems can efficiently supply energy for remote battery-limited sensors, with careful optimization required for the UAV's trajectory and positioning, as well as for network resource allocation. In addition to UAV-enabled WET, UAV-aided data collection was studied in~\cite{UAVDC1, UAVDC2, UAVDC3}, where the UAV's trajectory, user scheduling, and radio resource allocation were optimized to maximize delivery probabilities while minimizing its energy consumption. Furthermore, recent works looked into combining WET and WIT for UAV-enabled WPCNs, where the UAV serves as an energy emitter and a data collector. For instance, the authors in~\cite{UAVWETWIT1} investigated an UAV-enabled WPCN system, where both the time and position of the UAV are jointly optimized to maximize the uplink sum-rate for all users. The study~\cite{UAVWETWIT2} provided closed-form expressions for the energy outage probability and rate outage probability in UAV-enabled WPCNs, considering Rician fading channels and the UAV’s elevation angle. In~\cite{UAVWETWIT3}, the authors derived the optimal 3D-trajectory of multiple UAVs as well as the time allocation for WET and WIT within a limited time duration, aiming at maximizing the UAV’s worst data collection rate among all users. Additionally, the authors in~\cite{UAVWETTGCN} formulated the problem of maximizing the amount of offloaded data considering the propulsion energy consumption at the UAV, and derived a closed-form expression for the optimal time slots on different tasks for UAV-enabled WPCNs. However, these studies focus solely on terrestrial networks, leaving a gap in research on UAV-enabled WET and WIT for underground networks.

\subsection{Motivations and Contributions}
Our literature review reveals that the performance of UAV-enabled WET and WIT remains unclear for WPUCNs, where signal propagation is not only more lossy but is also affected by some new factors such as soil’s properties (e.g., texture and bulk density), volumetric water content (VWC), and burial depth. Furthermore, a thorough evaluation is lacking for UAV-enabled WET efficiency of the state-of-the-art CSI-free schemes. Currently, only one study investigated UAV-enabled WET in the underground domain, where a field experiment was conducted to demonstrate that a UD buried at a depth of $0.15$~m with a VWC of $40\%$ can harvest an average of $2$~dBm of RF energy at $915$~MHz from the PS with a transmit power of $36$~dBm. Note that the PS was suspended by the UAV at a height of $4$~cm above the ground~\cite{UAVWETField}. However, this study focused on the point-to-point communication and did not consider multi-antenna WET and WIT integration in massive WPUCNs. 

Motivated by these research gaps, in this work, we explore the potential feasibility of utilizing UAVs as both a PS and a data collector in a multi-user WPUCN system for large-scale underground monitoring. Specifically, the UAV is first wirelessly charged by the terrestrial hybrid access point (HAP), and then it flies over the monitoring area to charge the UDs followed by the sensor data collection from all UDs, and finally it returns to the HAP for data offloading. In contrast to~\cite{UAVWETTVT, UAVWETTGCN}, our study extends UAV-enabled WPCNs into the underground domain, employs CSI-free schemes, and emphasizes time allocation strategies for minimizing UAV energy consumption while satisfying the throughput requirement for each UD and ensuring complete sensor data offloading to the HAP. To the authors’ best knowledge, this work is the first to assess the feasibility and performance of combining WET and WIT in large-scale UAV-enabled WPUCN scenarios. The specific contributions of this study are summarized as follows:
\begin{itemize}
    \item We conceptualize a UAV-enabled WPUCN system that integrates energy transfer from the HAP to the UAV, the UAV's WET and WIT operations, and data offloading from the UAV to the HAP. We also establish an energy consumption model that characterizes the power budget during the WET phase and the UAV’s propulsion consumption. Furthermore, we propose a hybrid WET approach that enables UDs to harvest energy from both the HAP and the UAV, leveraging full-CSI and CSI-free multi-antenna WET. 

    \item We consider three different WET approaches: the traditional PS approach, where UDs harvest energy only from the PS; the UAV-enabled WET approach, where UDs harvest energy from the UAV; and our proposed hybrid approach. For each approach, we formulate and solve an optimization problem to minimize the UAV's energy consumption by appropriately scheduling time slots for various processes, including the UAV's charging by the HAP, UAV’s WET to UDs, UDs' WIT to the UAV, and data offloading to the HAP. The problem is subject to the time, UD throughput, and data offloading constraints.

    \item Through simulations of real-world farm scenarios, we demonstrate that the proposed hybrid approach outperforms  conventional WET approaches in terms of the average worst-case RF energy available at UDs. However, its performance gain depends on the number of antennas, communication distance, UDs' number, burial depth, and soil’s VWC. Moreover, we show that under the derived time allocation, the hybrid approach achieves the lowest UAV energy consumption while ensuring the throughput requirement for each UD when an appropriate CSI-free scheme and number of antennas are adopted for the HAP and UAV. Our results indicate that the derived time allocation based on the hybrid WET approach, can achieve efficient WET and WIT operations. This paves the way for the deployment of practical UAV-enabled WPUCNs for large-scale, cost-effective, and sustainable underground monitoring.
\end{itemize}

\subsection{Article Organization and Notations}
\begin{table}[!t]
\caption{List of Acronyms}
\label{Tablist}
\centering
\begin{tabular}{|m{0.07\textwidth}<{\raggedright}|m{0.36\textwidth}<{\raggedright}|}
\hline
\textbf{Acronym} & \textbf{Definition} \\
\hline
AAIS   & All antennas transmitting independent signals \\
\hline
AASS   & All antennas transmitting the same signal \\
\hline
CSI    & Channel state information \\
\hline
EH     & Energy harvesting \\
\hline
HAP    & Hybrid access point \\
\hline
LOS    & Line-of-sight \\
\hline
NLOS   & Non-line-of-sight \\
\hline
PS     & Power source \\
\hline
PWM    & Pulse with modulation \\
\hline
RF     & Radio-frequency \\
\hline
RAB    & Rotary antenna beamforming \\
\hline
SA     & Switching antennas \\
\hline
SDP    & Semidefinite programming \\
\hline
TDMA   & Time-division multiple access \\
\hline
UD     & Underground device \\
\hline
UAV    & Unmanned aerial vehicle \\
\hline
ULA    & Uniform linear array \\
\hline
VWC    & Volumetric water content \\
\hline
WET    & Wireless energy transfer \\
\hline
WIT    & Wireless information transfer \\
\hline
WUSN   & Wireless underground sensor network \\
\hline
WPCN   & Wireless powered communication network \\
\hline
WPUCN  & Wireless-powered underground communication network \\
\hline
\end{tabular}
\end{table}

The remainder of this article is organized as follows. Section~\ref{SystemSec} describes the system model. Section~\ref{WETSchemeSec} provides an illustration of the full-CSI and CSI-free schemes. Section~\ref{ECModelSec} models the power budget and UAV’s propulsion consumption. Section~\ref{TimeSec} introduces three WET approaches and formulates an optimization problem to minimize the UAV's energy consumption, and derives the optimal time allocation. Finally, Section~\ref{ResSec} presents and analyzes the simulation results, and Section~\ref{ConSec} concludes the article.

\textit{Notation:} Boldface lowercase and uppercase letters represent column vectors and matrices, respectively. For instance, $\mathbf{x}=\{x_i\}$, where $x_i$ is the $i$-th element of vector $\mathbf{x}$, while $\mathbf{X}=\{x_{i,j}\}$, where $x_{i,j}$ is the $i$-th row $j$-th column element of matrix $\mathbf{X}$. We denote a vector of ones by $\mathbf{1}$. Superscripts $(\cdot)^H$ and $(\cdot)^T$ represent the conjugate transpose and transpose operations, respectively, while the operator $\operatorname{Tr}(\cdot)$ denotes the trace. Furthermore, $\mathbbm{E}[\cdot]$ represents the statistical expectation, while $\inf\{\cdot\}$ is the infimum notation. Additionally, $\mathbb{C}$ and $\mathbb{R}$ denote the sets of complex and real numbers, respectively, and $\mathbbm{i} = \sqrt{-1}$ is the imaginary unit. The curled inequality symbol $\succeq$ is used to indicate positive definiteness of a matrix. Finally, $\mathbf{w} \sim \mathcal{CN}(\mathbf{0}, \mathbf{R})$ denotes a circularly symmetric complex Gaussian random vector with zero mean and covariance matrix $\mathbf{R}$. Table~\ref{Tablist} lists the abbreviations used throughout this article.

\section{System Model} \label{SystemSec}
\begin{figure}[!t]
    \centering
    \includegraphics[width=3.45in]{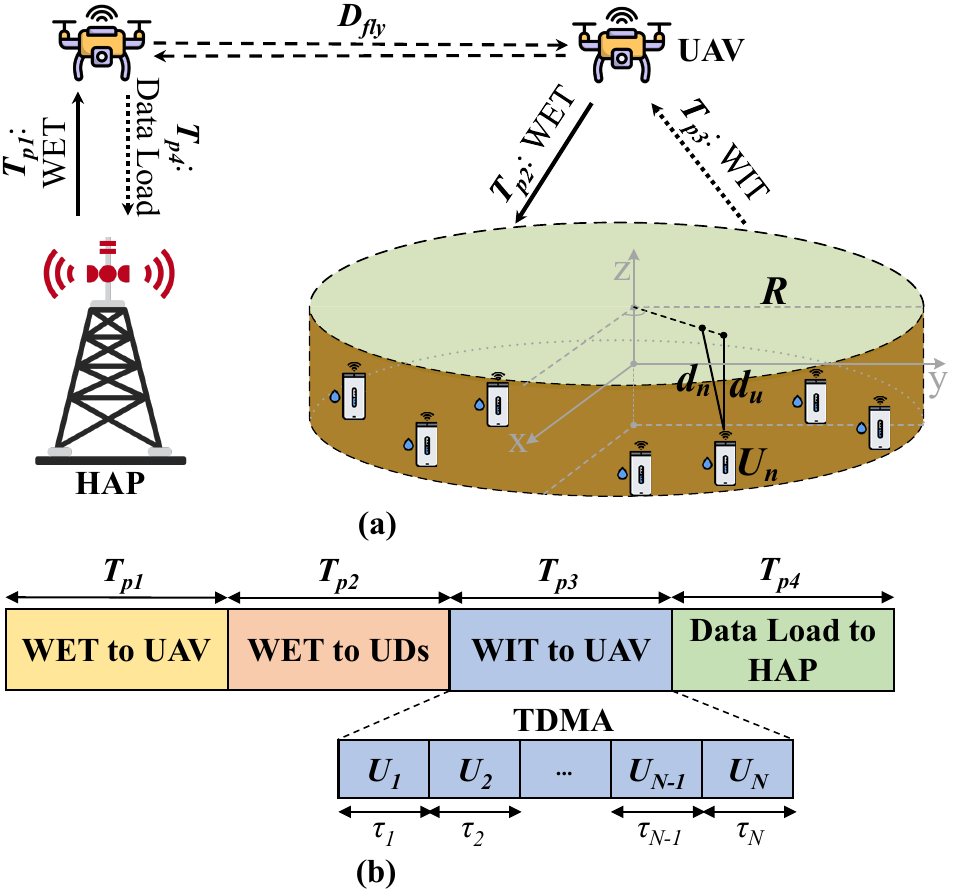}
    \caption{A UAV-enabled WPUCN system model. (a) System mode. (b) Time block}
    \label{Sysmod_fig}
\end{figure}

Consider a multi-user UAV-enabled WPUCN system. As depicted in Fig.~\ref{Sysmod_fig}(a), the system comprises one terrestrial HAP, one UAV equipped with a hybrid communication and power transfer module, and a set of single-antenna UDs, denoted by $\mathcal{U}=\{U_n|n=1, 2, \ldots, N\}$, distributed within a monitoring circle of radius $R$ with the same burial depth $d_u$. The HAP and the UAV support both WET and WIT function. Specifically, the HAP serves not only as a PS providing WET to both the UAV and UDs but also as a gateway for receiving the data from the UAV. Meanwhile, the UAV can perform WET to UDs and collect their data. Let $(0,0, H_{hap})$ denote the location of HAP at a height of $H_{hap}$, and the $n$-th UD is located at $(x_i, y_i, -d_u)$. The UAV is set to fly at a fixed altitude of $H_{uav}$ above the ground, with its location denoted by $(x, y, H_{uav})$. Herein, we assume that both the HAP and UAV are equipped with a uniform linear array (ULA) of $Q$ half-wavelength spaced antennas. 

The UAV-assisted data collection process in the WPUCN system are divided into four phases, as described in Fig.~\ref{Sysmod_fig}(b). In Phase 1, a rotary-wing UAV is wirelessly charged by the HAP in its proximity during the charging duration $T_{p1}$. Once charging is complete, the UAV flies a distance of $D_{fly}$ to the center of the monitoring area. In Phase 2, the UDs are charged using WET in the downlink during the duration $T_{p2}$. In Phase 3, the UDs employ the harvested energy to perform WIT to the UAV in the uplink following time-division multiple access (TDMA) protocols within the duration $T_{p3}$. Specifically, $U_n$ subsequently transmits sensed data during its allocated time duration $\tau_n$ to avoid inter-user interference among UDs. In Phase 4, the UAV flies back to the HAP and completely offloads the collected data during the duration $T_{p4}$.

Our goal in this work is to determine the optimal time allocation among these four phases, given a fixed flight distance $D_{fly}$, with the objective of minimizing the UAV's energy consumption while meeting the throughput threshold requirements of each UD. Note that the round-trip process is not included in the time allocation, as the flight time is determined by the distance $D_{fly}$ and the speed $V$ of UAV, both of which are constant~\footnote{Deriving the optimal UAV's speed that minimizes energy consumption during the round-trip flight, given the fixed distance $D_{fly}$, is an interesting direction and is left for future work.}. In this study, we assume that the multi-antenna technique is activated only in Phase 2 for enhancing WET efficiency, while remaining inactive during other phases to reduce power consumption and hardware complexity~\footnote{Adopting multi-antenna techniques in other phases entails proper beamforming design for single-input multiple-output channels in Phase 3, and multiple-input multiple-output channels in Phase 1 and 4, potentially improving system performance. However, such a design is not straightforward and the optimization must account for the extra power consumption, and it has been left for future work as it definitely constitutes an interesting and promising research direction.}. The details of the four phases are provided in the following subsection.

\subsection{Phase 1 -- Charging Phase}
In Phase I, the HAP charges the UAV for its maneuvering and WET operations within the duration $T_{p1}$. Recently, various technologies for UAV wireless charging have been developed, including capacitive power transfer, inductive power transfer, magnetic resonant coupling, laser beamforming, and RF power transmission~\cite{UAVChargeReview}. Among these, RF power transmission is chosen for wirelessly charging the UAV in Phase 1 due to its efficient long-range power delivery. Assume that the communication channel between the HAP and the UAV is dominated by LOS, the received energy at the UAV is given by 
\begin{equation}
    \label{EuavrEq}
    E_{uav-r} = \zeta  \frac{P_{hap} G_{hap} G_{uav} |h_0|^2}{\left(4 \pi f (H_{uav}-H_{hap}) /c\right)^2} T_{p1}, 
\end{equation}
where $\zeta \in [0,1)$ denotes the energy conversion efficiency, i.e., the efficiency of converting RF energy into direct current, $P_{hap}$ is the transmit power of the HAP, $G_{hap}$ and $G_{uav}$ are the antenna gains of the HAP and the UAV, respectively, $f$ is the carrier frequency, $c$ denotes the speed of light in free space, while $|h_0|^2$ is the small-scale fading coefficient between the HAP and the UAV. Herein, we assume the channel coefficients are perfectly known and/or vary very slowly for both uplink and downlink between the HAP and the UAV, thus, set $|h_0|^2=1$ without loss of generality~\cite{LinBSWPUSNs, H1ref}. 

\subsection{Phase 2 -- WET Phase}
In Phase 2, the UAV hovers at the center of the monitoring area with a height of of $H_{uav}$ and the PS  (either the HAP, the UAV, or both) charges all UDs via WET within $T_{p2}$. Herein, we consider a generic WET scenario where a PS equipped with a ULA of $Q$ antennas and wirelessly broadcasting RF energy to all UDs. Note that the PS can either be fixed at a specific location (e.g., HAP as shown in Fig.~\ref{Sysmod_fig}(a)) or deployed on a UAV to charge UDs that are far away.  

\subsubsection{Small-Scale Fading Model}
In the multi-antenna WET operation, we assume quasi-static channels, where fading remains constant over each transmission block and is independent and identically distributed (i.i.d.) across blocks. The channel experiences Rician fading, which can model a wide range of channel conditions by adjusting the Rician factor $\kappa$. For instance, the channel envelope follows a Rayleigh distribution when $\kappa=0$, while a LOS component is introduced for $\kappa>0$, getting stronger as $\kappa$ increases.~\cite[Ch. 2]{Ricianfading}. Accordingly, the normalized channel vector between the ULA of PS and $U_{n}$ is expressed as~\cite[Ch. 5]{Ricianeq}
\begin{equation}
    \label{HnEq}    
    \mathbf{h}_{n}\left(\theta_{n}\right) = \sqrt{\frac{\kappa}{1+\kappa}}\mathbf{h}_{n}^{\mathrm{los}}\left(\theta_{n}\right)+\sqrt{\frac{1}{1+\kappa}}\mathbf{h}^{\mathrm{nlos}},
\end{equation}
where $\mathbf{h}_{n}^{\mathrm{los}}\left(\theta_{n}\right)=e^{\mathbbm{i} \vartheta_{0}}\left[1, e^{\mathbbm{i} \phi_{1, n}}, e^{\mathbbm{i} \phi_{2, n}}, \ldots, e^{\mathbbm{i} \phi_{Q-1, n}}\right]^{\mathrm{T}}$ is the deterministic LOS component, while $\mathbf{h}_{\mathrm{nlos}}\sim\mathcal{CN}(\mathbf{0},\mathbf{R})$ accounts for the NLOS channel under the scattering (Rayleigh) fading. More specifically, $\vartheta_{0}$ is an initial phase shift, which can be ignored as it affects all antenna elements equally. Meanwhile, $\phi_{t, n}$, $t \in \{1, \ldots, Q-1\}$ represents the mean phase shift of the ($t+1$)-th antenna element relative to the first antenna element as observed by $U_n$, and is given by~\cite{CSIfreeIoT}. 
\begin{equation}
    \phi_{t, n} = -t \pi \sin(\theta_{n}),
\end{equation}
where $\theta_{n} \in [0, 2\pi]$ is the azimuth angle of $U_{n}$ relative to the transmitting ULA, which depends on both the PS's ULA orientation and the UD's location.

\subsubsection{Path Loss Model} 
The total path loss from the PS to $U_n$, i.e., $ \delta_{n}$, consists of the above-ground air attenuation $J_{n}$, the refraction loss at the air-soil interface $K_{n}^{a2u}$, and the attenuation in underground soil $M_{n}$. Herein, we adopt the modified Friis-based model developed in~\cite{LinLoRaWUSNs, AnnaWUSNs, Undergroundfield}, which has been validated through field experiments conducted at various depths and under different soil conditions, demonstrating accurate estimation of attenuation in soil. Mathematically, the model is expressed as
\begin{align}
 \label{EqTotalPL} \delta_{n} &= J_{n} K_{n}^{a2u} M_{n},\\
\label{airpathloss} J_{n} (l_n)&=\left(\frac{4 \pi f}{c}\right)^2 l_{n}^\varsigma, \\
\label{Rpathloss} K_{n}^{a2u}&=\left(\frac{\sqrt{\left(\sqrt{\varepsilon'^{2}+\varepsilon''^{2}}+\varepsilon' \right) / 2}+1 }{4}\right)^2, \\
\label{soilpathloss} M_{n}(d_n)&=\left(\frac{2 \beta d_{n}}{e^{-\alpha d_{n}}}\right)^{2},
\end{align}
where $\varsigma$ is the path-loss exponent, while $l_{n}$ and $d_{n}$ represent the propagation distances through air and underground soil, respectively, from the PS to $U_{n}$. Since the permittivity of soil is significantly higher than that of air, most of the RF signal energy from the above-ground source will be reflected if the incident angle is large. Therefore, we consider that the refracted angle is nearly zero during the RF signal propagation from air to underground soil~\cite{ Undergroundfield}. Thus, we assume in this study that the propagation in the soil is vertical, implying $d_{n}=d_{u}$. Additionally, $\alpha$ and $\beta$ denote the attenuation constant and phase shifting constant, respectively, which are given as
\begin{align}
\alpha &= 2 \pi f \sqrt{\frac{\mu_{r} \mu_{0} \varepsilon' \varepsilon_{0}} {2}\left[\sqrt{1+\left(\frac{\varepsilon''}{\varepsilon'}\right)^{2}}-1 \right]}, \\
\label{beta}\beta &= 2 \pi f \sqrt{\frac{\mu_{r} \mu_{0} \varepsilon' \varepsilon_{0}} {2}\left[\sqrt{1+\left(\frac{\varepsilon''}{\varepsilon'}\right)^{2}}+1 \right]}.
\end{align}
Herein, $\mu_{r}$ is the soil’s relative permeability, $\mu_{0}$ is the free-space permeability, $\varepsilon_{0}$ is the free space permittivity, and $\varepsilon'$ and $\varepsilon''$ are the real and imaginary parts of the soil’s relative permittivity, respectively, i.e., $\varepsilon = \varepsilon' + j\varepsilon''$. Notice that $\varepsilon$ can be calculated by the mineralogy-based soil dielectric model~\cite{MBSDM}. MBSDM can operate over a wide frequency range, from $45$~MHz to $26.5$~GHz, and provides accurate predictions of $\varepsilon$ as it is derived from a large number of soil samples and accounts for the presence of both free and bound water in the soil. It requires only three input parameters to compute the complex permittivity of soil: the VWC, the operating frequency of the RF signals, and the clay percentage of the soil.

\subsubsection{Average Received Power} 
When the PS transmits $K \le Q$ energy symbols $\{s_{k}\}$ per channel, the received RF signal $y_{n}$ at $U_{n}$ can be modeled by 
\begin{equation}
\label{signal}
    y_{n}=\sum_{k=1}^K \sqrt{\frac{p_{k}}{\delta_{n}}} \mathbf{h}_{n}^{\mathrm{T}} \mathbf{v}_k s_{k},
\end{equation}
where $\mathbf{v}_k = [v_{k}^{(1)}, v_{k}^{(2)},\ldots, v_{k}^{(Q)}] \in \mathbb{C}^Q$ is the normalized precoding vector associated with $s_{k}$, it depends on the selected WET scheme, either full-CSI or CSI-free, while $p_{k}$ is the transmit power corresponding to each energy symbol and $\sum_{k=1}^{K}p_{k} = p$. The energy symbols $\{s_k\}$ are assumed to be i.i.d. unit-power and zero-mean random variables, i.e.,  $\mathbb{E}[|s_{k}|^2]=1$, $ \mathbb{E}[s_k]=0,$ and $\mathbb{E}[s_{k}^H s_{k^\prime}]=0~\forall k \neq k^\prime$. Consequently, the incident RF power (averaged over the signal waveform) at $U_{n}$ is given by 
\begin{align}
\label{pmi}
\xi_{n} &=\mathbb{E}_{s_k}\left[\left|y_{n}\right|^2\right] \nonumber \\
& \stackrel{(a)}{=} \mathbb{E}_{s_k}\!\! \!\left[\!\!\left(\sum_{k=1}^K \! \sqrt{\frac{p_{k}}{\delta_{n}}} \mathbf{h}_{n}^{\mathrm{T}} \mathbf{v}_k s_{k}\!\!\right)^{\mathrm{\! \! \! H}} \!\!\! \left(\sum_{k=1}^K \! \sqrt{\frac{p_{k}}{\delta_{n}}} \mathbf{h}_{n}^{\mathrm{T}}  \mathbf{v}_ks_{k}\!\!\right)\!\!\right] \nonumber \\
& \stackrel{(b)}{=} \frac{1}{\delta_{n}} \sum_{k^{\prime}=1}^K \sum_{k^{\prime \prime}=1}^K \!\! \sqrt{p_{k^{\prime}} p_{k^{\prime \prime}}}\!\left(\mathbf{h}_{n}^{\mathrm{T}} \mathbf{v}_{k^{\prime}}\right)^{\!\mathrm{H}} \!\!\mathbf{h}_{n}^{\mathrm{T}} \mathbf{v}_{k^{\prime \prime}} \mathbb{E}\left[s_{k^{\prime}}^{\mathrm{H}} s_{k^{\prime \prime}}\right] \nonumber \\
& \stackrel{(c)}{=} \frac{1}{\delta_{n}} \sum_{k=1}^K p_{k}\left|\mathbf{h}_{n}^{\mathrm{T}} \mathbf{v}_k\right|^2,
\end{align}
where (a) comes from leveraging~\eqref{signal}, (b) follows after reorganizing terms, and (c) is obtained based on the assumption of i.i.d. power-normalized signals.

In a typical quasi-static WET setup, the energy harvested by $U_{n}$ under a linear energy harvesting (EH) model is directly proportional to the average incident RF power, as expressed by
\begin{equation}
    \label{Eneq}
    E_n = \zeta G_{ps}G_{ud}\xi_{n},
\end{equation}
where $G_{ps}$ and $G_{ud}$ are the antenna gains of the PS and the UDs, respectively. Although nonlinear EH models are intrinsically more accurate due to the nonlinearities of the EH hardware, the harvested power benefits from an increased average incident RF power either under a linear or nonlinear EH model~\cite{PracticalEH}. For the sake of simplicity, we focus on the linear EH model and defer the analysis and related discussions of the nonlinear EH model to future work.

\subsection{Phase 3 -- WIT Phase}
After all UDs are charged, they utilize the harvested energy to transmit sensor data to the UAV during its allocated time slots $\tau_n$, with $n=1, \ldots, N$, using TDMA protocol in the uplink within the duration $T_{p3}$. This implies that $T_{p3} = \sum_{n=1}^{N} \tau_n$, as illustrated in Fig.~\ref{Sysmod_fig}(b). The total path loss from $U_n$ to the UAV consists of the underground path loss $M_n$, the air attenuation $J_n$, and the refraction loss from soil to air $K_n^{u2a}$.  Consequently, the achievable throughput of $U_n$ during $\tau_n$ in Phase 3 can be expressed as 
\begin{equation}
    R_{n} = \tau_n W \log_2\left(1+\frac{\varphi E_{n} G_{ud} G_{uav} |h_{n}^{ud}|^2}{\tau_n M_n K_n^{u2a} J_n \sigma_{A}^{2} }\right),
\end{equation}
where $W$ is the channel bandwidth, $\sigma_{A}^{2}$ denotes the variance of the additive white Gaussian noise, the channel coefficients from the UDs to the UAV are assumed to be $|h_{n}^{ud}|^2=1, \forall n$ for simplicity, while $\varphi$ represents the portion of received energy used for WIT. Note that the remaining ($1-\varphi$) portion of the harvested energy supports the circuit operations. Furthermore, since most energy is refracted when the signal propagates from soil to air, the refraction loss on the soil-air interface can be neglected, implying $K_n^{u2a}=1$~\cite{Undergroundfield, LinBSWPUSNs}. 

After Phase 3, the sum-data received by the UAV from $N$ UDs can be calculated by
\begin{equation}
    R_{uav} = \sum_{n=1}^{N} R_n. \label{RuavEq}
\end{equation}

\subsection{Phase 4 -- Data Loading Phase}
In Phase 4, the UAV returns to the HAP and hovers above it for data offloading. The amount of offloaded data at the HAP during the duration $T_{p4}$ is given by 
\begin{align}
    \label{RhapEq}
    R_{hap} = &T_{p4} W \log_{2} \left(1+\frac{P_{uav} G_{uav} G_{hap} |h_0|^2}{\left(4 \pi f (H_{uav}-H_{hap})/c\right)^2 \sigma_{A}^{2}}\right), 
\end{align}
where $P_{uav}$ is the transmit power of the UAV. Furthermore, the sensor data from all UDs should be completely loaded to the HAP, implying $R_{hap} = R_{uav}$.

\section{Wireless Energy Transfer Schemes} \label{WETSchemeSec}
In Phase 2, the efficiency of multi-antenna WET can be enhanced by designing an appropriate precoding scheme. In the following Subsection~\ref{FullCSISec}, we present a full-CSI precoding scheme for optimizing the WET process. Then, we introduce several state-of-the-art CSI-free WET alternatives in Subsection~\ref{CSIfreeSec} to efficiently powering all UDs without any CSI acquisition.

\subsection{Full-CSI schemes} \label{FullCSISec}
In the full-CSI scheme, the PS transmits pilot signals, which are used by the UDs to estimate the CSI for the downlink channels. Once this information is fed back to the PS, it uses the estimated CSI to optimize the precoder, ensuring maximum fairness in charging the UDs. This implies that no UD is expected to benefit more than others from the PS’s WET. To further tilt the scale in favor of the CSI-free schemes, we assume an ideally trained full-CSI strategy, disregarding the impact of imperfect CSI and the time and power consumed for both CSI acquisition and the precoder optimization.\footnote{Notice that assessing the power consumed in the CSI acquisition phase is not straightforward as it may depend on several system parameters that are not considered in this work, e.g., pilot transmit power, number of pilot symbols, and multiuser pilot scheduling. Furthermore, the effects of imperfect CSI should also be taken into account. Consequently, we have assumed an ideally trained full-CSI strategy, which may somewhat bias the results in its favor.} This assumption aligns with the setup in~\cite{RAB, LinCSIfree, OnelWCL}.

We define $\xi_{csi}$ as the minimum harvested energy among all UDs under the full-CSI scheme, i.e., $\xi_{csi} \triangleq \inf_{n=1, \ldots, N}\{\xi_n\}$, where $\xi_n = \frac{p}{\delta_n} \operatorname{Tr}(\mathbf{V} \mathbf{H}_n)$ is reformulated from Eq.~\eqref{pmi} with $\mathbf{V} = \sum_{k=1}^{K} \mathbf{v}_k \mathbf{v}_k^{H}$ and $\mathbf{H}_n = \mathbf{h}_n \mathbf{h}_n^H$. Given $\mathbf{V}$ is a Hermitian matrix with a maximum rank of $Q$, the optimization problem can be formulated as a semidefinite programming (SDP) problem, i.e.,
\begin{subequations}
\begin{align} 
\label{P1}({\rm{P1}}):{\mathop {\min} \limits_{{\mathbf{V} \in \mathbb{C}^{Q \times Q},~ \xi_{csi}}}} &{-\xi_{csi}}\\  
s.t.~ \label{c1_1} &\frac{p}{\delta_{n}} \operatorname{Tr}\left(\mathbf{V} \mathbf{H}_n\right) \geq \xi_{csi}, \forall n,\\
      \label{c1_2} &\operatorname{Tr}\left(\mathbf{V}\right)=1,\\
      \label{c1_3} &\mathbf{V} \succeq 0.
\end{align}
\end{subequations}
The CVX toolbox can efficiently solve this SDP problem and obtain the eigenvectors of $\mathbf{V}$, which is referred to as the optimal full-CSI beamforming~\cite{OnelWCL}.

\subsection{State-of-the-Art CSI-free Schemes}\label{CSIfreeSec}
\begin{figure}[!t]
    \centering
    \includegraphics[width=3.45in]{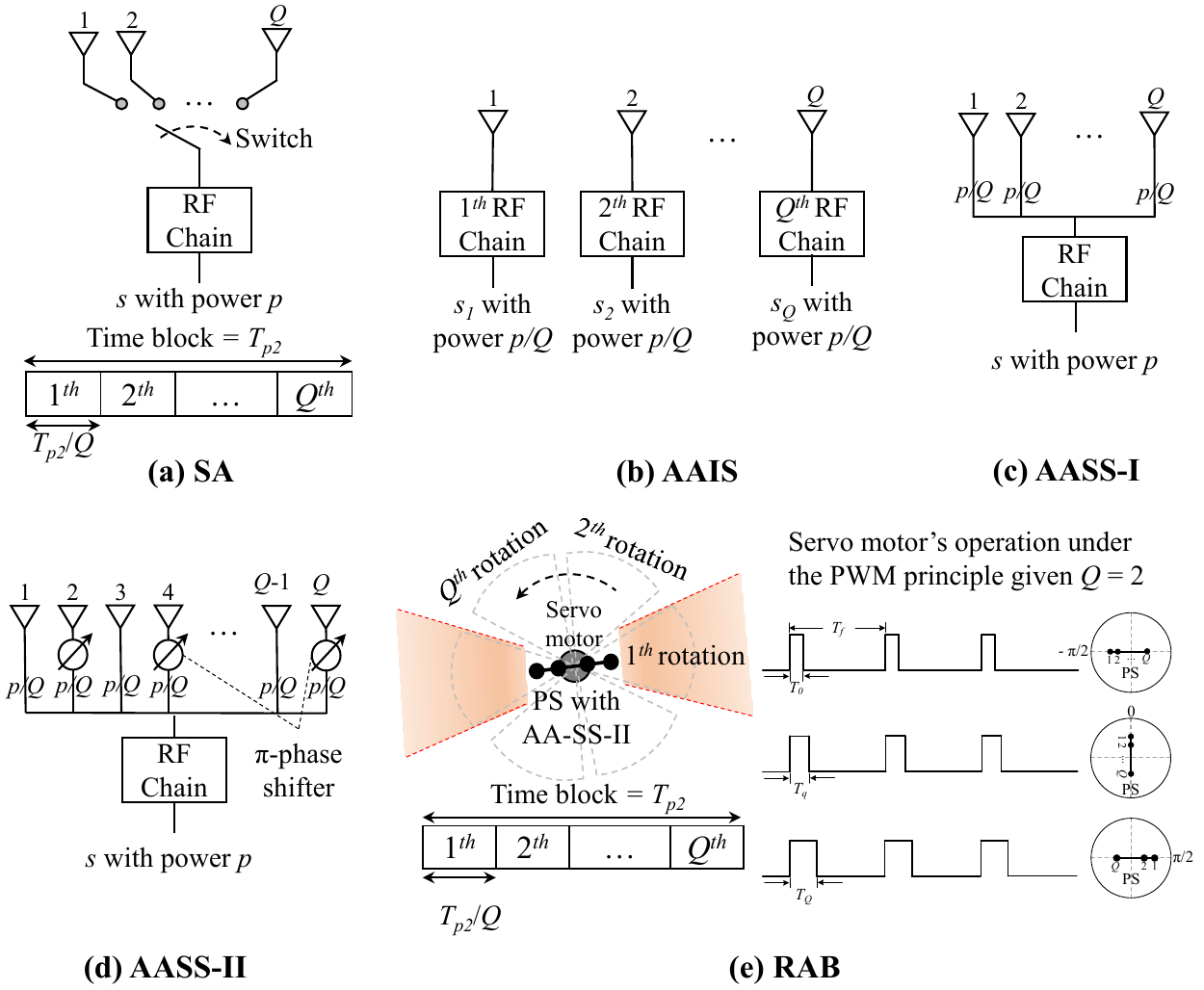}
    \caption{The operation diagram of (a) SA, (b) AAIS, (c) AASS-I, (d) AASS-II, and (e) RAB.}
    \label{CSIfreeOpFig}
\end{figure}

\begin{figure}[!t]
    \centering
    \includegraphics[width=3.45in]{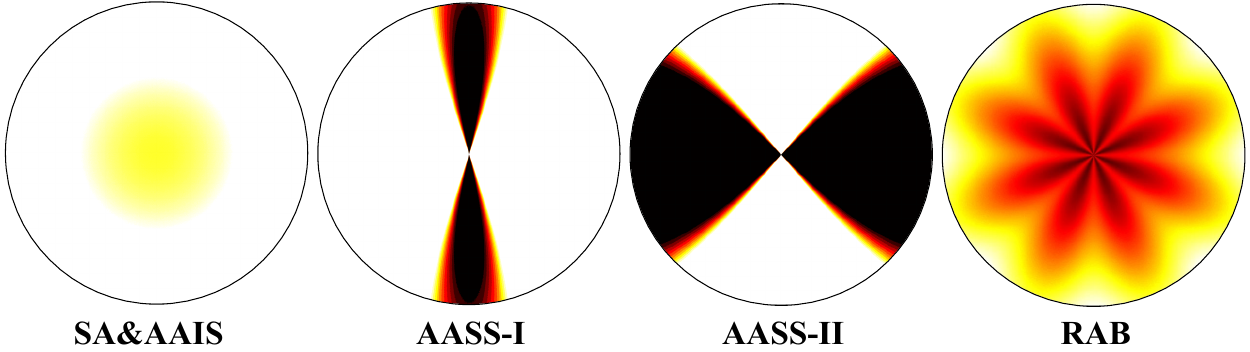}
    \caption{The radiation pattern of SA, AAIS, AASS-I, AASS-II, and RAB.}
    \label{RaditionFig}
\end{figure}

Although the PS employs the full-CSI scheme to enhance WET efficiency while ensuring fairness among UDs, reliable and accurate CSI acquisition is challenging and costly, and even infeasible in massive WPUCN scenarios due to the harsh underground soil~\cite{LinCSIfree}. Moreover, the power consumed for CSI acquisition and the SDP-based solution can eliminate the benefits gained from CSI exploitation~\cite{OnelWETReview, LinLoRaWUSNs}. By intelligently exploiting the broadcast nature of wireless transmissions, several promising CSI-free schemes are proposed to efficiently charge a large set of nearby UDs. Compared to the full-CSI scheme, CSI-free schemes avoid the need for CSI acquisition and may offer reliable WET performance even in complex underground environments; however, their WET efficiency remains inherently affected by underground conditions~\cite{LinCSIfree}. Herein, we briefly explain the state-of-the-art CSI-free schemes that will be adopted and comparatively analyzed for Phase 2 in our UAV-enabled WPUCN system. Figs.~\ref{CSIfreeOpFig} and~\ref{RaditionFig} respectively illustrate the implementations and radiation patterns of SA, AAIS, AASS, and RAB. 

\subsubsection{SA~\cite[Sec. III-A]{CSIfreeTcom}}
Under SA as depicted in Fig.~\ref{CSIfreeOpFig}(a), the PS utilizes a switching mechanism to transmit a signal with the full power of the $q$-th antenna during the $q$-th duration such that $Q$ antennas are used over the WET phase $T_{p2}$. We assume equal time allocation among antennas such that each subblock duration is equivalent to $T_{p2}/Q$. Since only one antenna is active at the $q$-th subblock duration, SA requires a single RF channel for its operation, implying $K=1$, $p_k=p$, and $\mathbf{v}_{k}$ is a one-dimensional column vector containing the scalar $1$ in Eq.~\eqref{pmi}. Note that the total incident RF energy is the sum of the energy from $Q$ subblocks.

\subsubsection{AAIS~\cite[Sec. III-B]{CSIfreeIoT}} 
Instead of transmitting a signal with one antenna at a time, the PS using AAIS transmits signals independently generated across the antenna elements and with equal transmit power, thus $K=Q$ and $p_k = p/Q$, as highlighted in Fig.~\ref{CSIfreeOpFig}(b). Therefore, in Eq.~\eqref{pmi}, $v_{k}^{(q)}=1$ for $k=q$, otherwise $v_{k}^{(q)}=0$. Different from SA, $Q$ RF chains are required to implement AAIS since all antenna elements are simultaneously active to transmit $Q$ independent RF signals. However, it is evidenced in~\cite{CSIfreeIoT} that SA has equal/similar WET performance to AAIS under a linear/nonlinear EH model. Furthermore, the radiation patterns for both SA and AAIS are omnidirectional, as shown in Fig.~\ref{RaditionFig}.

\subsubsection{AASS} In AASS, the same signal is transmitted through all antenna elements with equal power, i.e., $K=1$ and $p_k = p/Q$ in Eq.~\eqref{pmi}. There are two configurations for AASS:
\begin{enumerate}[itemindent=1em]
    \item AASS-I~\cite[Sec. III-A]{CSIfreeTcom}, where the precoding vector $\mathbf{v}_{k}=\mathbf{1}$ in Eq.~\eqref{pmi}, or simply no precoder, is applied to attain an energy beam towards the ULA's boresight directions, as shown in Fig.~\ref{CSIfreeOpFig}(c).
    
    \item AASS-II~\cite[Sec. IV-B]{CSIfreeIoT}, where the precoding vector is set as $v_{k}^{(q)}=e^{\mathrm{mod}(q-1,2)\pi \mathbbm{i}}$ in Eq.~\eqref{pmi} to attain wider energy beams, which are offset $90^\circ$ from ULA's boresight directions. This can be realized with an analog implementation with a number of $\lfloor Q/2\rfloor$ $\pi$-phase shifters, as displayed in Fig.~\ref{CSIfreeOpFig}(d).
\end{enumerate}
The gains of both AASS schemes are strongly associated with the UDs' positions and the orientation of the PS's ULA; therefore, they are preferable when charging UDs clustered in specific boresight directions. The radiation patterns of AASS-I and AASS-II are directed towards the ULA boresight and a $90^\circ$ offset from ULA boresight directions, respectively, as highlighted in Fig.~\ref{RaditionFig}.

\subsubsection{RAB~\cite{RAB}} 
As exhibited in Fig.~\ref{CSIfreeOpFig}(e), the servo motor is equipped in the PS to continuously rotate its antenna array while adopting the AASS-II scheme, which allows improving the charging coverage probability. By taking advantage of the symmetry of the ULA’s radiation patterns, we consider $Q$ angular rotations to cover the angular domains $[-\pi/2, \pi/2]$. Indeed, a servo motor can rotate the antenna array at specific angles and realize $Q$ equally spaced steps using the pulse with modulation (PWM) technique, that can provide sufficiently smooth performance. Note that the $q$-th rotation step is conducted during the $q$-th subblock duration, where each duration is $T_{p2}/Q$. According to Eq.~\eqref{pmi}, the incident average RF power gain at $U_{n}$ under RAB is given by
\begin{align}
\label{RABenergy}
\xi_{n}^{RAB}= &\frac{1}{Q} \sum_{q=1}^Q  \sum_{k=1}^K \frac{p_{k}}{\delta_{n}} \left|\mathbf{h}_{n}^{\mathrm{T}}\mathbf{v}_k\left(\theta_{n}+\frac{q \pi}{Q}\right)\right|^2,
\end{align}
where $\theta_{n}$ is the initial azimuth angle prior to any rotation. The resulting radiation pattern is quasi-omnidirectional, as shown in Fig.~\ref{RaditionFig}. Note that the operation of RAB requires at least $Q=2$ antenna elements; otherwise, it is equivalent to AASS-II.

\section{Energy Consumption Model}\label{ECModelSec}
In this section, we examine the power budget for the PS to carry out the full WET operations in Phase 2, as well as the UAV's propulsion consumption for hovering and round-trip flight, within the energy consumption model of the proposed UAV-enabled WPUCN system.

\subsection{Power Budget Model}
Due to the power consumed by the transmitter's power amplifier, circuitry, and operations, the budgeted power $P_b$ is not entirely converted to transmit power $p$ in practice for the full-CSI and CSI-free schemes in Phase 2. Herein, we consider the impact of the circuitry, base-band operations and the servo motor rotation of RAB on the power consumption. Meanwhile, we assume that the full-CSI scheme is implemented in fully digital ULAs without considering the power consumed by both the CSI acquisition and the SDP-based solution. Note that $Q$ RF chains are required in the full-CSI scheme and AAIS implementations, whereas the other CSI-free schemes only employ one RF chain. Compared to the power consumption in RF chains, the power consumed by the switch (in the case of SA) and phase shifters (in the case of AASS-II and RAB) is negligible in this study~\cite{Switchpower, Phasepower, LinCSIfree}. Furthermore, RAB requires the servo motor to continuously rotate its antenna array. Therefore, compared to the other CSI-free schemes, the extra power consumed by the servo motor operations should be carefully considered in RAB. Accordingly, the transmit power of the PS can be calculated by~\cite{LinCSIfree}
\begin{equation}
    \label{p_pra}
    p = \eta (P_{b}-Q^i P_{rf}-P_c-bP_{motor}),
\end{equation}
where $\eta$ is the amplifier efficiency, $P_b = P_{hap}$ when the HAP serves as the PS and $P_b = P_{uav}$ when the UAV operates as the PS, $P_{rf}$ is the power consumed by the base-band processing per RF chain, $P_c$ is the fixed power consumption considering the remaining circuitry, and $i=1$ for the full-CSI scheme and AAIS, and $i=0$ for the other CSI-free schemes. Here, $P_c$ and $P_{rf}$ are assumed to be constant without loss of generality. Furthermore, $b=1$ for RAB, and $b=0$ for the other WET schemes. By using the PWM principle, the servo motor can rotate an antenna array with $Q$ equally spaced steps in the angular domains $[-\pi/2,\pi/2]$, as illustrated in Fig.~\ref{CSIfreeOpFig}(e). Considering the practical implementation of servo motor rotations, the power consumed by the servo motor operations, i.e., $P_{motor}$, is given by~\cite{LinCSIfree}
\begin{equation}
    P_{motor} = \frac{\sum_{q=0}^{Q}{T_0+\frac{q}{Q}}}{T_f} V_{motor} I_{motor},
\end{equation}
where $T_0$ is the pulse width for the shaft at the initial angle before the rotation operations, $T_f$ is the duty cycle, $V_{motor}$ is the supply voltage of the servo motor, while $I_{motor}$ is the working current during rotation.

\subsection{UAV's Propulsion Consumption Model}
In addition to the energy charged to the UDs, the UAV also requires energy for various maneuvering operations, such as hovering, acceleration, deceleration, and flying at constant speed. An analytical propulsion power consumption model for rotary-wing UAVs flying at speed $V$ was proposed in~\cite{UAVEnergyModel}, which is given by 
\begin{equation}
\label{UAVEnergyEq}
P(V) = P_0\!\!\left(1+\frac{3 V^2}{U_{tip}^2}\right)\!+\!P_i\sqrt{\sqrt{1+\frac{V^4}{4 v_0^4}}-\frac{V^2}{2 v_0^2}}\!+\!\frac{d_0 \rho S A V^3}{2},
\end{equation}
where $P_0$ and $P_i$ are two constants related to the physical properties of UAV and the flight environments, such as weight, rotor radius and air density, $U_{tip}$ represents the tip speed of the rotor blade, $v_0$ is the mean rotor induced velocity in hover, $d_0$ and $S$ are the fuselage drag ratio and rotor solidity, respectively, while $\rho$ and $A$ denote the air density and rotor disc area, respectively. 

By substituting $V=0$ into Eq.~\ref{UAVEnergyEq}, we obtain the power consumption for hovering status, i.e., $P(0) = P_0+P_i$. Hence, the energy required for hovering can be expressed as 
\begin{equation}
    E_{h}(t_h) = P(0) t_h, 
\end{equation}
where $t_h$ is the hovering time. Meanwhile, the energy consumption for UAVs flying at speed $V$ is given by  
\begin{equation}
    E_{V} = P(V) T_{fly-V},
\end{equation}
where $T_{fly-V}$ is the time for the UAV flying at a constant speed of $V$. In this study, we consider that the UAV accelerates from an initial velocity of 0 to $V$, then continues to fly towards the UDs at a constant speed of $V$, and finally decelerates from $V$ to 0 to hover over the center of monitoring area. Hence, we can obtain $T_{fly-V}=\frac{D_{fly}}{V}-\frac{V}{a}$ and the flying time of the UAV $T_{fly}=\frac{D_{fly}}{V}+\frac{V}{a}$ with the acceleration $a$, where $T_{fly}$ will be used in Eq.~\eqref{Eq_EnPS} to determine the received energy of $U_n$. Since acceleration and deceleration are symmetric in this process, the energy consumed during deceleration is the same as that during acceleration. Consequently, the energy consumed for the UAV during the acceleration and deceleration can be calculated by~\cite{UAVWETTVT, UAVWETTGCN}  
\begin{equation}
    E_{acc} = E_{dec} = \int_{0}^{\frac{V}{a}} P(t) dt,
\end{equation}
where $P(t) = P_0\left(1+\frac{3 (at)^2}{U_{t i p}^2}\right)+P_i\sqrt{\sqrt{1+\frac{(at)^4}{4 v_0^4}}-\frac{(atr)^2}{2 v_0^2}} +\frac{d_0 \rho s A (at)^3}{2}$ by substituting $V=at$ into Eq.~\eqref{UAVEnergyEq}.

As illustrated in Fig.~\ref{Sysmod_fig}, the UAV hovers above the HAP to be charged wirelessly, then accelerates to a constant speed $V$ and flies towards the UDs. Upon approaching the center of the monitoring area, it decelerates from $V$ to $0$ and hovers above the UDs to perform WET and collect sensor data before returning to the HAP in the same way. Finally, the UAV hovers above the HAP for data offloading. Therefore, the total energy consumption of the UAV excluding the UAV's charging process can be expressed as 
\begin{equation}
    \label{EsEq}
    E_{s} = E_{wet}+E_{dl}+E_{h}(T_{p2})+E_{h}(T_{p3})+ E_{h}(T_{p4})+E_{ft}+E_{fb},
\end{equation}
where $E_{wet} = P_{uav} T_{p2}$ and $E_{dl} = P_{uav} T_{p4}$ denote the energy required for WET and data loading operations, respectively, $E_{h}(T_{p2})$,  $E_{h}(T_{p3})$, and $E_{h}(T_{p4})$ respectively represent the energy consumed for hovering during WET, WIT, and data loading operations, while $E_{ft}$ and $E_{fb}$ correspond to the energy required for the round-trip flight, given by $E_{ft} = E_{fb} = E_{acc} + E_v + E_{dec}$.

\section{Time Allocation for UAV-enabled WPUCNs} \label{TimeSec}
In Subsection~\ref{WETAPPSec}, we first present three WET approaches for the WET Phase. We formulate the energy consumption minimization problem and derive the optimal time allocation in Subsection~\ref{TimesolveSec}. 

\subsection{WET Approaches} \label{WETAPPSec}
\begin{figure}[!t]
    \centering
    \includegraphics[width=3.45in]{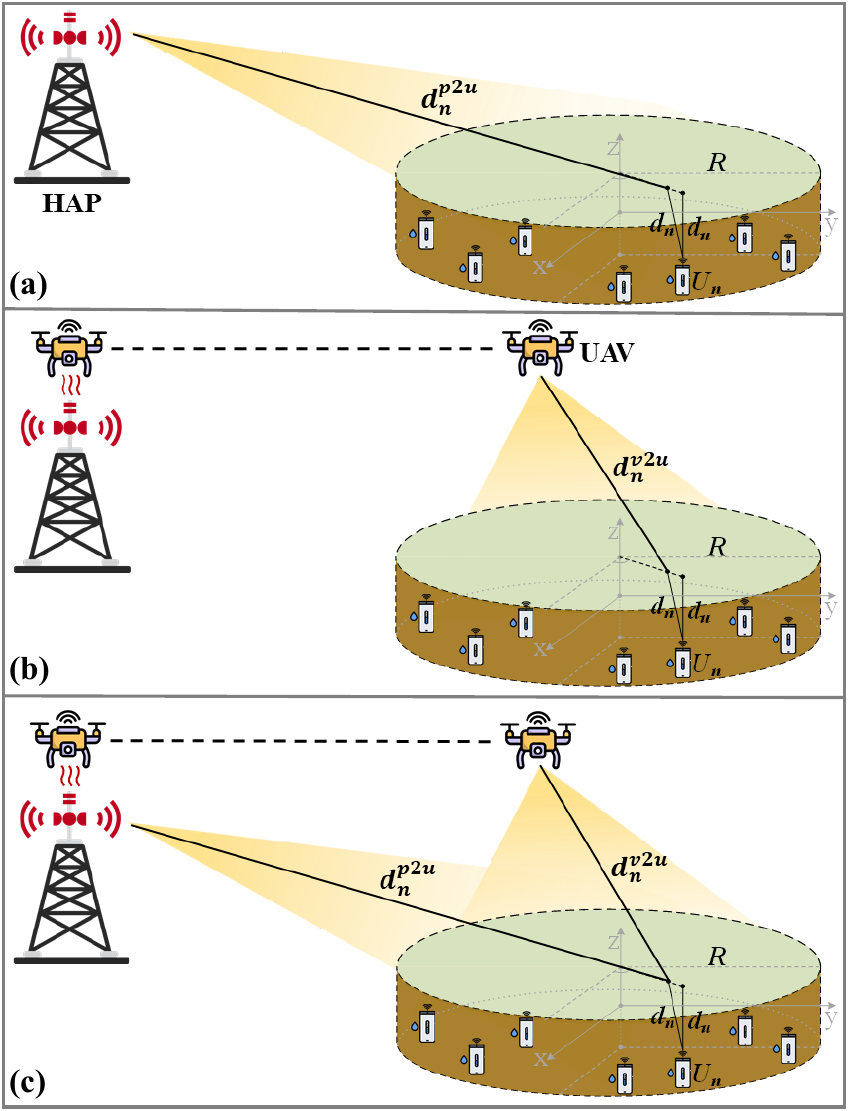}
    \caption{Three WET approaches for WPUCN system: (a) traditional PS approach, (b) UAV-enabled WET approach, and (c) hybrid approach.}
    \label{UAVWETArch_fig}
\end{figure}

In Phase 2, three WET approaches are considered to charge all UDs, as illustrated in Fig.~\ref{UAVWETArch_fig}.
\subsubsection{Traditional PS Approach~\cite{LinCSIfree}}  \label{PSSec}
As shown in Fig.~\ref{UAVWETArch_fig}(a), the HAP acts as the PS to wirelessly charge the UDs within the duration $T_{p2}$ and the UAV‘s flying time $T_{fly}$. Herein, either full-CSI and CSI-free WET schemes are adopted to enable WET operations. In this case, the UAV is used solely for data collection, without the WET functionality. By leveraging Eq.~\eqref{Eneq}, the received energy at $U_n$ can be expressed as
\begin{equation}
    \label{Eq_EnPS}
    E_{n}^{ps} = \zeta (T_{fly}+T_{p2}) G_{hap} G_{ud} \xi_{n}\left(\kappa_{p2u},\varsigma_{p2u}, P_{hap}, d_{n}^{p2u}\right),
\end{equation}
where $\kappa_{p2u}$ and $\varsigma_{p2u}$ are the Rician factor and path loss exponent for the HAP-to-UDs channels, respectively, while $d_n^{p2u}$ is the air propagation length between the HAP and $U_n$. These parameters are used in Eq.~\eqref{HnEq} and Eq.~\eqref{EqTotalPL} to calculate the received energy of $U_n$ from the HAP.

\subsubsection{UAV-enabled WET Approach~\cite{UAVWETTVT}} \label{UAVWETSec}
In the UAV-enabled WET approach as illustrated in Fig.~\ref{UAVWETArch_fig}(b), the ULA-equipped UAV hovers at an altitude $H_{uav}$ above the center of the monitoring area and serves as the PS to charge all UDs in the downlink during $T_{p2}$ with either full-CSI or CSI-free WET schemes. In this phase, the HAP remains silent and does not provide the WET functionality. Based on Eq.~\eqref{EqTotalPL}, the received energy at $U_n$ under this approach is given by 
\begin{equation}
    E_{n}^{uav} = \zeta T_{p2}  G_{uav} G_{ud} \xi_{n}\left(\kappa_{v2u},\varsigma_{v2u}, P_{uav}, d_{n}^{v2u}\right),
\end{equation}
where $\kappa_{v2u}$ and $\varsigma_{v2u}$ are the Rician factor and path loss exponent for UAV-to-UDs channels, respectively, while $d_n^{p2u}$ is the air propagation length between the UAV and $U_n$. These parameters are used in Eq.~\eqref{HnEq} and Eq.~\eqref{EqTotalPL} to obtain the received energy of $U_n$ from the UAV. 

\subsubsection{Hybrid Approach} \label{HybridSec}
As illustrated in Fig.~\ref{UAVWETArch_fig}(c), the proposed hybrid approach utilizes both the HAP and UAV as PSs to broadcast RF energy to all UDs, employing either full-CSI or CSI-free WET schemes. Ignore the energy contributed by coexisting/neighboring networks operating in the same spectrum and assume that the HAP and UAV generate and transmit independent signals, the received energy at $U_n$ can be calculated as
\begin{equation}
   E_{n}^{hybrid} = E_{n}^{ps}+E_{n}^{uav}.
\end{equation}
Since the UDs can harvest energy from both the HAP and UAV, the hybrid approach is expected to deliver more efficient WET compared to traditional PS and UAV-enabled approaches. In Phase 2, we assume that both the HAP and UAV use the same WET scheme, which could be either full-CSI or CSI-free. It is important to select an appropriate CSI-free scheme for the HAP and UAV, respectively, within the hybrid approach to ensure the optimal WET performance. 

\subsection{Time Allocation Optimization} \label{TimesolveSec}
Based on these WET approaches in Phase 2, our goal to minimize the UAV's energy consumption in all phases, while ensuring the throughput of each UD and complete data offloading to the HAP. This translates to finding the optimal time allocation (i.e., $T_{p2}$, $T_{p3}=\sum_{n=1}^N \tau_n$, and $T_{p4}$ as illustrated in Fig.~\ref{Sysmod_fig}(b)) that minimizes the energy consumption of the UAV, i.e., $E_s$, by considering the constraints on time, the throughput requirements for each UD, and data offloading. The throughput-aware energy consumption minimization problem for the considered UAV-enabled WPUCN system is formulated specifically as
\begin{subequations}
\begin{align} 
\label{P2}({\rm{P2}}):{\mathop {\min} \limits_{{T_{p2}}, {T_{p3}}, {T_{p4}}}} &{E_{s}}\\  
\operatorname{s.t.}~ \label{c2_1} &T_{p2}>0,\\
      \label{c2_2} &\tau_n>0, \forall n,\\
      \label{c2_3} &T_{p4}>0,\\
      \label{c2_4} &R_{n} \geq \gamma_{n}, \forall n,\\
      \label{c2_5} &R_{hap} = R_{uav},
\end{align}
\end{subequations}
where Eq.~\eqref{c2_1}, ~\eqref{c2_2}, and~\eqref{c2_3} correspond to the time constraints, Eq.~\eqref{c2_4} defines the throughput requirements for each UD, implying that the achievable throughput of each UD should exceed the throughput threshold $\gamma_n$, while Eq.~\eqref{c2_5} ensures that all sensor data from the UDs are offloaded to the HAP.

Since the UAV's energy consumption is directly proportional to $T_{p2}$, $T_{p3}$, and $T_{p4}$, and $T_{p4}$ is determined when $R_{hap}=R_{uav}$ as defined in Eq.~\eqref{RuavEq} and~\eqref{RhapEq}, the optimization problem (P2) can be reformulated as
\begin{subequations}
\begin{align} 
\label{P3}({\rm{P3}}):&{\mathop {\min} \limits_{{T_{p2}}, \tau_1, \ldots, \tau_N}} {T_{p2}}+\sum_{n=1}^N \tau_n\\ 
\operatorname{s.t.}~\label{c3_1} &T_{p2}>0,\\
      \label{c3_2} &\tau_n>0, \forall n\\
      \label{c3_3} &R_{n} \geq \gamma_{n}, \forall n.\\
      \label{c3_4} &T_{p4} = \frac{R_{uav}}{W \log_{2} \left(1+\frac{P_{uav} G_{uav} G_{hap}}{\left(4 \pi f (H_{uav}-H_{hap})/c\right)^2 \sigma_{A}^{2}}\right)}.
\end{align}
\end{subequations}
We prove that (P3) is a convex problem. First, the objective function, Eq.~\eqref{c3_1} and~\eqref{c3_2} are linear, thus convex. Related to~\eqref{c3_3}, let's substitute proceed writing
\begin{align}
    R_n = \tau_n W \log_2 \left({1+\frac{C_n T_{p2} + b_n}{\tau_n}}\right).
\end{align}
Herein, for the traditional PS approach, $C_n$ and $b_n$ are respectively given by 
\begin{align}
    C_n &= \frac{ \varphi G_{ud} G_{uav} \zeta \xi_{n}\left(\kappa_{p2u},\varsigma_{p2u}, P_{hap}, d_{n}^{p2u}\right)}{\delta_{n} \sigma_{A}^{2}}, \\
    b_n &= \frac{ \varphi G_{ud} G_{uav} \zeta T_{fly} \xi_{n}\left(\kappa_{p2u},\varsigma_{p2u}, P_{hap}, d_{n}^{p2u}\right)}{\delta_{n} \sigma_{A}^{2}}.
\end{align}
For the UAV-enabled WET approach, $C_n$ and $b_n$ are respectively given by
\begin{align}
    C_n &= \frac{ \varphi G_{ud} G_{uav} \zeta \xi_{n}\left(\kappa_{v2u},\varsigma_{v2u}, P_{uav}, d_{n}^{v2u}\right)}{\delta_{n} \sigma_{A}^{2}}, \\
    b_n &= 0.
\end{align}
For the proposed hybrid approach, $C_n$ and $b_n$ are respectively given by
\begin{align}
    C_n &=\frac{\varphi G_{ud} G_{uav} \zeta }{\delta_{n} \sigma_{A}^{2}} \left(\xi_{n}\left(\kappa_{v2u},\varsigma_{v2u}, P_{uav}, d_{n}^{v2u}\right)+\right. \\ \nonumber
    &~~~~\left.\xi_{n}\left(\kappa_{p2u},\varsigma_{p2u}, P_{hap}, d_{n}^{p2u}\right)\right), \\
    b_n &= \frac{ \varphi G_{ud} G_{uav} \zeta T_{fly} \xi_{n}\left(\kappa_{p2u},\varsigma_{p2u}, P_{hap}, d_{n}^{p2u}\right)}{\delta_{n} \sigma_{A}^{2}}.
\end{align}

Accordingly, constraint~\eqref{c3_3} can be rewritten as 
\begin{equation}
    f(\tau_n, T_{p2}) = \frac{\tau_n}{C_n} \left(2^{\frac{\gamma_n}{\tau_n W}}-1\right)-\frac{b_n}{C_n}-T_{p2} \leq 0.
\end{equation}
Next, we examine the convexity of the constraint by analyzing the Hessian matrix of the $f(\tau_n, T_{p2})$ for all WET approaches as described in Section~\ref{PSSec},~\ref{UAVWETSec}, and~\ref{HybridSec}. The Hessian matrix for all WET approaches is identical and is given by
\begin{equation}
\mathbf{Z}_n = \begin{bmatrix}
\frac{\gamma_n^2 (\ln 2)^2}{C_n W^2} \frac{1}{\tau_n^3} 2^{\frac{\gamma_n}{W \tau_n}} & 0\\
0  & 0
\end{bmatrix}.
\end{equation}
Obviously, this Hessian matrix is positive semidefinite, implying $\mathbf{Z}_n \succeq 0$. Therefore, the constraint~\eqref{c3_3} is convex. Since both the objective function and the constraints are convex, (P3) is a convex optimization problem~\cite{convexbook}. Consequently, the optimization toolbox (e.g., CVX)~\cite{cvx} can be adapted to effectively solve this problem. Note that problem (P3) is always feasible for any positive throughput requirement under valid channel conditions, since one can always adjust $T_{p2}$ or $T_{p3}=\sum_{n=1}^N \tau_n$ to ensure  $\gamma_n$. The feasibility of (P3) is primarily determined by the throughput threshold $\gamma_n$, the channel gain parameters (i.e., $C_n$ and $b_n$), and the system bandwidth $W$.

After determining the optimal values of $T_{p2}$, $T_{p3}=\sum_{n=1}^{N} \tau_n$, and $T_{p4}$, we can obtain the total energy consumption of the UAV $E_s$ in Eq.~\eqref{EsEq}. To ensure that the energy harvested by the UAV from the HAP is sufficient to support all operations during Phases 2, 3, and 4, it must hold that $E_{uav-r} \geq E_s$, where $E_{uav-r}$ is given by Eq.~\eqref{EuavrEq}. Since the UAV consumes power for hovering operation while being charged, the minimum required time for the UAV to be charged by the HAP, i.e., $T_{p1}$, can be calculated using Eq.~\eqref{EuavrEq} and~\eqref{EsEq} as follow
\begin{equation}
\label{Tp1Eq}
    T_{p1} = \frac{E_{s}}{\zeta  \frac{P_{hap} G_{hap} G_{uav}}{\left(4 \pi f (H_{uav}-H_{hap}) /c\right)^2}+P(0)}. 
\end{equation}
Therefore, the UAV is guaranteed to have sufficient energy to complete all required tasks as long as its charging time exceeds the derived minimum $T_{p1}$. The proposed UAV-enabled WPUCN system implements based on the derived time allocation strategy to collect the sensor data from all UDs with the minimum UAV's energy consumption.

\section{Numerical Results} \label{ResSec}
This section illustrates the WET efficiency of our proposed WET approaches considering both full-CSI and CSI-free schemes and the performance of our proposed  time allocation results under different throughput thresholds. To evaluate the practical performance of the UAV-enabled WPUCN system, we consider a real-world center-pivot irrigation farm as the study scenario. Unless stated otherwise, $N=64$ UDs are uniformly and randomly deployed within a circular area with a $5$~m radius and are buried at a depth of $0.4$~m with a VWC of $15\%$, while the HAP is positioned at a horizontal distance of $D_{fly}=600$~m from the center of the monitoring area. Note that these parameters should be adjusted based on the practical requirements and regional conditions of smart agriculture, thus we investigate the impact of varying these parameters on system performance in Section~\ref{WETEffResSec}. The \textit{in-situ} clay percentage of soil is obtained from \cite{Undergroundfield} to accurately estimate attenuation in soil. Both the HAP and the UAV are equipped with a ULA of $Q=32$ antennas, positioned at $4.5$~m and $5.5$~m above the ground, respectively. We set $\varsigma_{p2u} = 2.4$ and $\kappa_{p2u} = 3$ for the HAP-to-UDs channels, and  $\varsigma_{v2u} = 2$ and $\kappa_{v2u} = 10$ for the UAV-to-UDs channels in Phase 2. The system operates at a frequency band of $433$~MHz, which is ideal for underground wireless communications. The transmit power levels for the PB and the UAV are $35.56$~dBW and $10$~dBW, respectively. We define an EH threshold, $\psi$, such that the UDs can harvest energy only when the incident RF energy exceeds $\psi$. The power consumption parameters for the full-CSI and CSI-free schemes considering the transmitter power amplifier, circuitry, baseband operations, and servo motor rotation in RAB are summarized in Table~\ref{SimTab}~\cite{LinCSIfree}. Additionally, the UAV's propulsion consumption parameters are listed in Table~\ref{SimTab}~\cite{UAVWETTVT}.

We first assess the WET efficiency under various WET approaches in Section~\ref{CSIfreeResSec} and~\ref{WETEffResSec}, where a fixed UAV's charging time of $T_{p1}=120$~s is set to ensure sufficient energy for sustaining the UAV's operation and to obtain valid results even under challenging environmental conditions. Subsequently, in Section~\ref{TimeResSec}, we analyze time allocation strategies across different throughput requirements and WET approaches. Notably, the minimum required charging time $T_{p1}$ is determined using Eq.~\eqref{Tp1Eq}, based on the optimal time allocation obtained by solving (P3).

\begin{table}[t]
\caption{Simulation Parameters~\cite{LinCSIfree, UAVWETTVT}} 
\centering
\begin{tabular}{m{0.3\textwidth}<{\raggedright} m{0.15\textwidth}<{\centering}}%
\hline
\textbf{Parameters}                & \textbf{Values}                    \\  \hline
\multicolumn{2}{l}{\textbf{Operation Environments}}\\ \hline
Deployment Radius ($R$)            & 5~m                                \\    
Total number of nodes ($N$)        & 64                                 \\
UDs' deployment                    & uniform and random                 \\
Burial depth ($d_{u}$)             & 0.4~m                              \\
Number of antennas ($Q$)           & 32                                  \\
VWC ($m_v$)                        & 0.15                               \\
Clay ($m_c$)                       & 38\%                               \\
Carrier center frequency ($f$)     & 433~MHz                            \\
Channel bandwidth ($W$)            & 125~kHz                        \\
Transmit power of HAP and UAV ($P_{hap}$, $P_{uav}$)       &35.56 dBW, 10~dBW                   \\
Antenna gain of HAP, UAV, and UDs ($G_{hap}$, $G_{uav}$, $G_{ud}$)     & 15~dBi, 5~dBi, 2.15~dBi            \\
Height of HAP and UAV ($H_{hap}$, $H_{uav}$)              &4.5~m, 5.5~m                       \\
Distance from HAP to the center of monitoring area ($D_{fly}$) &600~m                \\
Path-loss exponents of HAP-to-UDs and UAV-to-UDs channels ($\varsigma_{p2u}$, $\varsigma_{v2u}$)  &2.4, 2 \\ 
Rician factor of HAP-to-UDs and UAV-to-UDs channels ($\kappa_{p2u}$, $\kappa_{v2u}$)           & 3, 10          \\ 
\hline
\multicolumn{2}{l}{\textbf{WET Configurations}}\\ \hline
Energy conversion efficiency ($\zeta$)       & 60\%                      \\
Portion of energy used for WIT ($\varphi$)& 60\%               \\
Amplifier efficiency ($\eta$)      & 38\%                               \\
Circuit power ($P_c$)              & 0.1~W                              \\
RF base-band consumption power ($P_{rf}$) & 0.06~W                      \\
Motor's duty cycle ($T_{f}$)       & 20~ms                              \\
Motor's voltage ($V_{motor}$)     & 5~V                                \\
Motor's current ($I_{motor}$)       & 250~mA                             \\
EH threshold  ($\psi$)           & -22~dBm                            \\ 
\hline
\multicolumn{2}{l}{\textbf{UAV's Energy Model Parameters}}\\ \hline
Flight speed of the UAV ($V$)            & 10~m/s                             \\
Acceleration/Deceleration ($a$)           &1~m/$s^2$                          \\
Charging time ($T_{p1}$)              &120~s                              \\
Bladed power ($P_0$)                &14.7517                            \\
Induced power ($P_i$)               &41.5409                            \\
Tip speed of the rotor blade ($U_{tip}$) &80                            \\
Mean rotor induced velocity  ($v_o$)&5.0463                             \\    Fuselage drag ratio ($d_0$)         &0.5009                             \\
Air density ($\rho$)                     &1.225~$kg/m^3$               \\
Rotor solidity ($S$)                &0.1248                      \\
Rotor disc area ($A$)               &0.1256~$m^2$                      \\
\hline
\label{SimTab}
\end{tabular}
\end{table}

\subsection{Performance of CSI-free Schemes} \label{CSIfreeResSec}
\begin{figure*}[!t]
    \centering
    \includegraphics[width=7in]{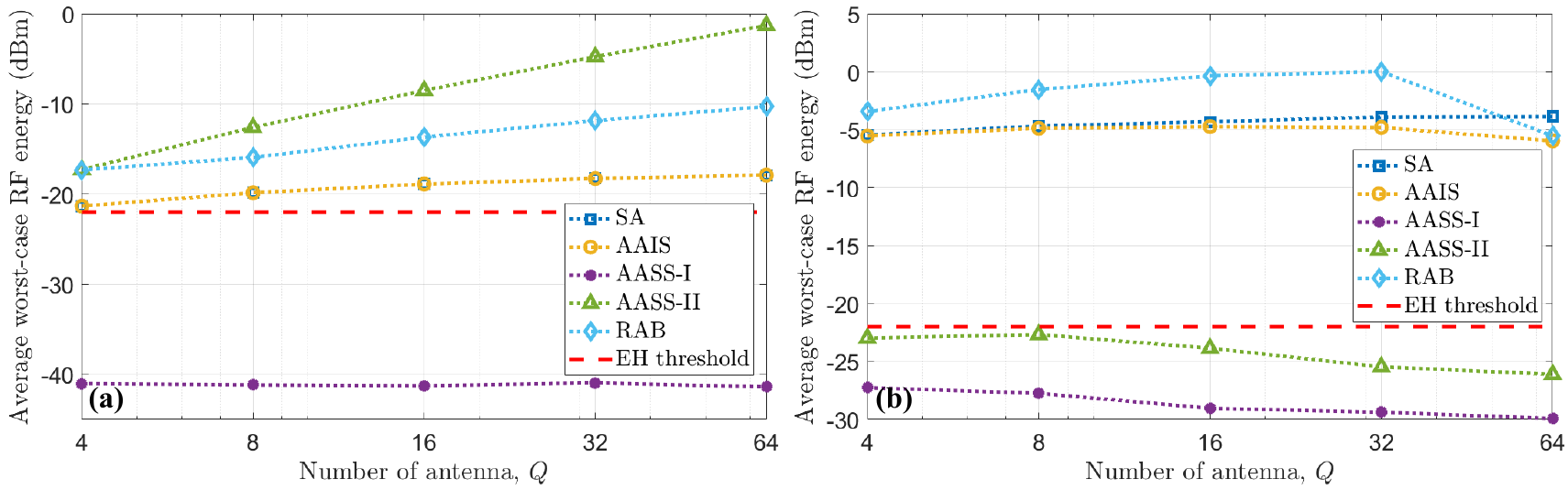}
    \caption{Average worst-case RF energy available for various CSI-free schemes as a function of the number of antenna $Q$ under (a) traditional PS and (b) UAV-enabled WET approaches, where the UAV charging time by the HAP is set to $T_{p1}=120$~s. The red dashed–dotted line depicts the EH threshold of $\psi = -22$~dBm.}
    \label{CSIfreeRes_fig}
\end{figure*}

We first present the average worst-case RF energy available at the input of the EH circuit across all UDs for the discussed CSI-free schemes, considering both the traditional PS and UAV-enabled WET approaches, as depicted in Fig.~\ref{CSIfreeRes_fig}. 

As illustrated in Fig.~\ref{CSIfreeRes_fig}(a), for the traditional PS approach, the average worst-case RF energy of SA, AAIS, AASS-II, and RAB increases with the number of antennas, while that of AASS-I remains nearly constant. Furthermore, AASS-II outperforms other CSI-free schemes, since the larger number of antennas leads to narrower energy beams at a $90^\circ$ offset from the ULA boresight directions, which corresponds to the monitoring area. Fig.~\ref{CSIfreeRes_fig}(b) shows that, for the UAV-enabled WET approach under RAB, the average RF energy increases with the number of antennas for $Q \leq 32$, and decreases from there on. This behavior is attributed to the fact that a larger number of antennas enhances the radiation performance of RAB, while simultaneously increasing the power consumption of the motor’s operations. Similarly, the performance of AAIS deteriorates when the number of antennas exceeds $32$, since the increase in RF chains causes higher baseband processing power consumption. When $Q \leq 32$, RAB exhibits better performance than other CSI-free UAV-enabled WET approaches. 

These findings highlight that the optimal WET performance can be achieved when the HAP employs the AASS-II scheme and the UAV utilizes the RAB scheme. Therefore, in the following results, when adopting the CSI-free scheme, the AASS-II scheme is applied to the traditional PS approach, while the RAB scheme is used in the UAV-enabled WET approach. Additionally, for our proposed hybrid approach, when utilizing the CSI-free scheme, we apply the AASS-II scheme to the HAP and the RAB scheme to the UAV.

\subsection{WET Efficiency Analysis}\label{WETEffResSec}
Next, we delve into the performance comparison among different WET approaches, considering both full-CSI and CSI-free schemes. Herein, the HAP adopts the AASS-II scheme, while the UAV utilizes the RAB scheme within the hybrid approach under the CSI-free scheme.

\begin{figure*}[!t]
    \centering
    \includegraphics[width=7in]{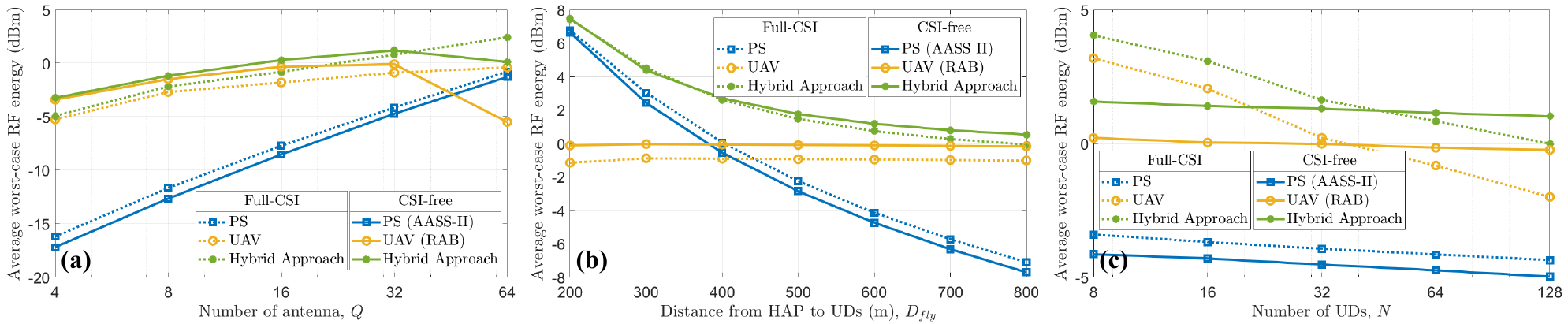}
    \caption{Average worst case RF energy available under various WET approaches with full-CSI and CSI-free schemes as a function of (a) the number of antennas $Q$, (b) the distance between the HAP and the center of the monitoring area, and (c) the number of UDs $N$, where the UAV charging time by the HAP is set to $T_{p1}=120$~s.}
    \label{WETRes1_fig}
\end{figure*}

\begin{figure*}[!t]
    \centering
    \includegraphics[width=7in]{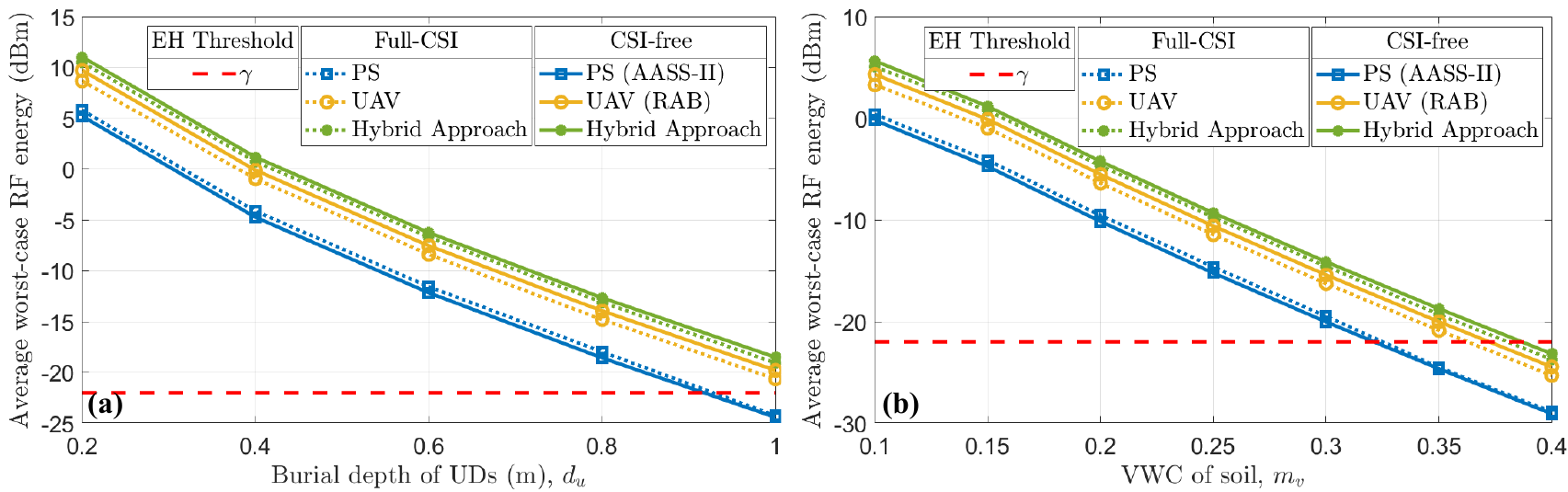}
    \caption{Average worst-case RF energy available under various WET approaches with full-CSI and CSI-free schemes as a function of (a) the burial depth of UDs $d_u$, and (b) the VWC of soil $m_v$, where the UAV charging time by the HAP is set to $T_{p1}=120$~s. The red dashed–dotted line depicts the EH threshold of $\psi = -22$~dBm.}
    \label{WETRes2_fig}
\end{figure*}

Fig.~\ref{WETRes1_fig} and~\ref{WETRes2_fig} depict the average worst-case RF energy availability delivered to the UDs for various WET approaches considering full-CSI and CSI-free schemes, under different conditions, including the number of antennas, flying distance, number of UDs, burial depths, and VWC. To achieve the optimal WET efficiency for the hybrid approach with the CSI-free schemes, the HAP employs the AASS-II scheme, while the UAV utilizes the RAB scheme. 
Fig.~\ref{WETRes1_fig}(a) shows that the average worst-case RF energy for the traditional PS approach, under both full-CSI and CSI-free schemes, increases with the number of antennas, same as the UAV-enabled and hybrid WET approaches with the full-CSI scheme. In contrast, the performance of the UAV-enabled WET and hybrid approaches with the CSI-free scheme improves for $Q \leq 32$ before deteriorating. This behavior is due to the limited power budget of the UAV and the power consumed by the circuitry and operations. For the traditional PS approach, the full-CSI scheme performs better than the CSI-free schemes, owing to the beamforming gain. In contrast, RAB outperforms the full-CSI scheme for UAV-enabled WET and hybrid approaches. This is because RAB exploits the mechanical rotation to improve the charging coverage probability, while the full-CSI scheme does not. When $Q \geq 32$, the hybrid approach with the CSI-free schemes outperforms other WET approaches, while the hybrid approach with the full-CSI scheme achieves the highest average RF energy at $Q = 64$. Note that the full-CSI scheme necessitates significant energy consumption for the CSI acquisition and precoding design via SDP.

Fig.~\ref{WETRes1_fig}(b) illustrates that the performance of the traditional PS and hybrid approaches deteriorates with the larger distance between the HAP and UDs due to the reduced energy harvested by the UDs from the HAP. The average worst-case RF energy for the UAV-enabled WET approach remains nearly constant, as the UDs harvest energy solely from the UAV hovering at the center of the monitoring area. Notably, the proposed hybrid approach exhibits the best performance among the WET approaches as the distance varies from $200$~m to $800$~m.    

Fig.~\ref{WETRes1_fig}(c) shows that the performance of all WET approaches deteriorates as the number of UDs increases from $8$ to $128$ due to the higher probability of UDs being farther from the HAP and the UAV. Note that the performance decline for the full-CSI scheme and AASS-II is more pronounced than for RAB. Such a phenomenon is due to the fact that the energy beams for the full-CSI scheme and AASS-II become less capable of efficiently reaching the UDs with the increasing number of UDs. Furthermore, the hybrid approach with the full-CSI scheme provides the highest RF energy when $8 \leq N \leq 32$, while the hybrid approach with the CSI-free schemes is the winner among all WET approaches as $64 \leq N \leq 128$.

As shown in Fig.~\ref{WETRes2_fig}(a), the average worst-case RF energy for all WET approaches decreases with the higher burial depth due to the increased attenuation from the longer propagation path through the underground soil. For instance, as the burial depth increases from $0.2$~m to $1$~m, the average worst-case RF energy decreases by approximately $30$~dBm. At the burial depth of $1$~m with the VWC of $15\%$, the traditional PS approach fails to exceed the EH threshold and are not capable of charging UDs, while the proposed hybrid approach with the CSI-free schemes can attain $-18.51$~dBm. Therefore, the proposed hybrid approach with the CSI-free schemes demonstrates the best performance, enabling efficient WET for UDs even under challenging underground conditions.

As illustrated in Fig.~\ref{WETRes2_fig}(b), the performance of all WET approaches deteriorates with VWC, as a higher VWC significantly increases attenuation in soil, which in turn greatly affects the incident RF energy. For instance, as the VWC increases from $0.1$ to $0.4$, the average worst-case RF energy of the proposed hybrid approach with the CSI-free schemes decreases from $5.63$~dBm to $-23.18$~dBm. For $m_v=0.4$ and $d_u=0.4$~m, none of the WET approaches surpass the EH threshold. To prevent unnecessary power expenditure in high-VWC underground conditions, the PS should halt the WET operation until the soil becomes drier, where the local VWC can be detected in real-time by UDs equipped with soil moisture sensors.

\subsection{Time Allocation Results} \label{TimeResSec} 
Finally, we illustrate the performance of our proposed time allocation scheme across various throughput thresholds and present the time allocation results considering different WET approaches.

\begin{figure}[!t]
    \centering
    \includegraphics[width=3.4in]{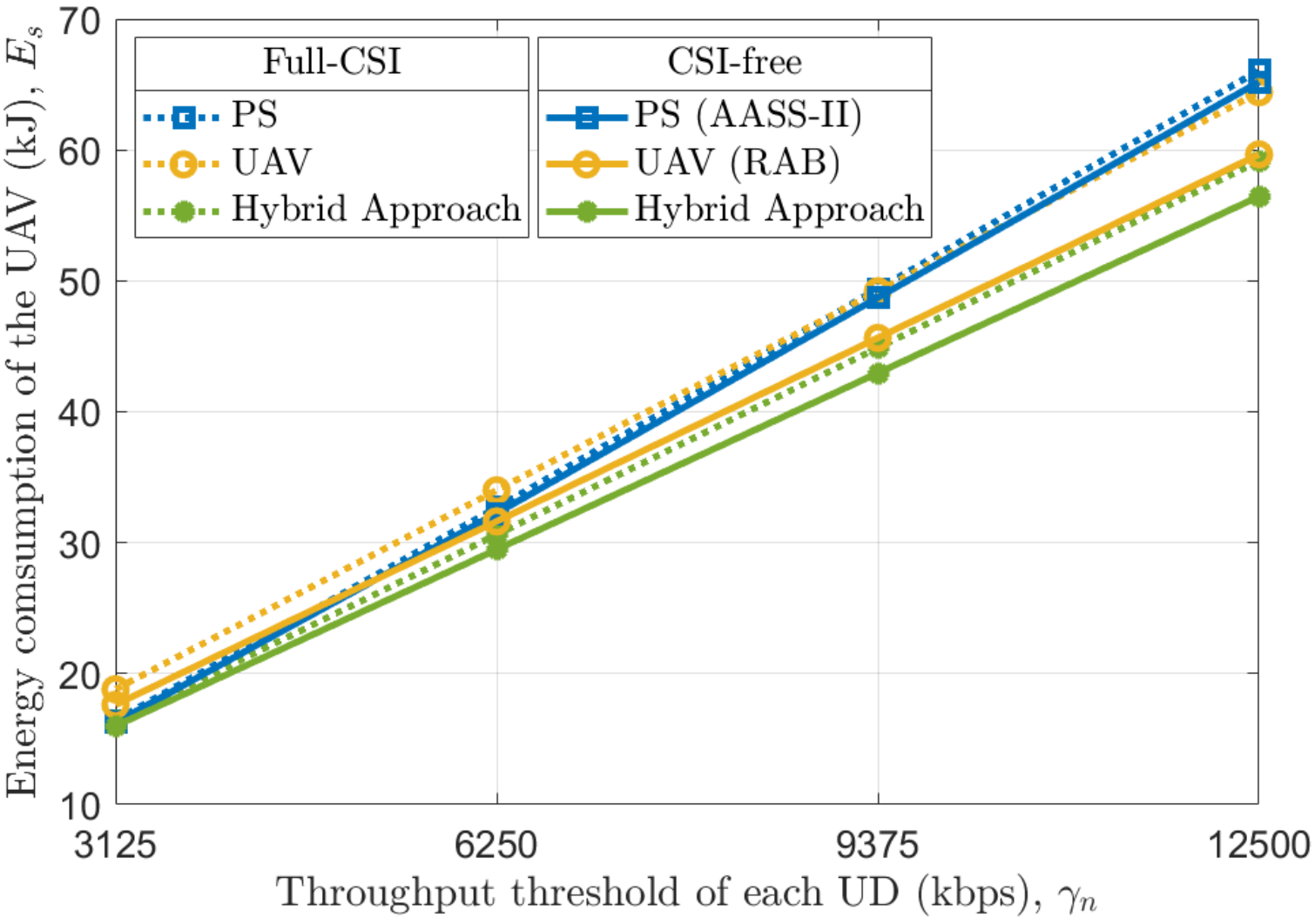}
    \caption{Total energy consumption of the UAV $E_s$ in the proposed UAV-enabled WPUCN system as a function of throughput thresholds $\gamma_n$ under the optimal time allocation considering various WET approaches with the full-CSI and CSI-free schemes.}
    \label{TimeRes_fig}
\end{figure}

Fig.~\ref{TimeRes_fig} describes the energy consumption of the UAV in all phases based on the optimal time allocation obtained by solving (P3), for different throughput thresholds $\gamma_n$ under various WET approaches with full-CSI and CSI-free schemes. Herein, we assume the same throughput thresholds for all UDs. As expected, the UAV's energy consumption increases with higher throughput thresholds as more time is required for UAV's WET and hovering operations. Note that our proposed hybrid approach outperforms all other WET approaches across all throughput thresholds for both full-CSI and CSI-free schemes. For instance, under the full-CSI scheme at $\gamma_n = 125000$~kbps, the UAV’s energy consumption under the hybrid approach decreases by $8.15\%$ and $10.74\%$ compared to the traditional PS and UAV-enabled WET approaches, respectively. Similarly, under the CSI-free scheme, the UAV's energy consumption for the hybrid approach is reduced by $13.43\%$ and $5.38\%$ compared to the traditional PS and UAV-enabled WET approaches at $\gamma_n = 125000$~kbps. Note that the hybrid approach utilizing the CSI-free scheme exhibits the lowest $E_s$ compared to all other WET mechanisms.

Table~\ref{CompareRes} summarizes the time allocation results obtained by solving (P3) with a throughput threshold of $\gamma_n = 12500$~kbps, considering various WET approaches under the full-CSI and CSI-free schemes, where $T_{total} = T_{p1}+T_{p2}+T_{p3}+T_{p4}$. The proposed hybrid approach outperforms other WET approaches. For instance, under the CSI-free scheme, the hybrid approach reduces the total time by $57$~s and $166$~s compared to the traditional PS and UAV-enabled WET approaches, respectively. For the hybrid approach, the total time of the CSI-free scheme is approximately $51$~s less than that of the full-CSI scheme. This indicates that the proposed hybrid approach enables efficient WET operation without the need for CSI acquisition. Furthermore, the hybrid approach under the CSI-free scheme achieves the lowest energy consumption of the UAV among all WET approaches, with $E_s=56.45$~kJ.

\begin{table}[t]
\caption{Comparison of Time Allocation Results under Different WET Approaches with $\gamma_{n}^{th}=12500$~kbps}
\centering
\label{CompareRes}
\begin{tabular}{m{0.13\textwidth}<{\raggedright} m{0.03\textwidth}<{\centering} m{0.03\textwidth}<{\centering} m{0.03\textwidth}<{\centering} m{0.03\textwidth}<{\centering} m{0.04\textwidth}<{\centering} m{0.03\textwidth}<{\centering}}
\toprule
\textbf{Approach} &$T_{p1}$ (s) &$T_{p2}$ (s) &$T_{p3}$ (s) &$T_{p4}$ (s) &$T_{total}$ (s) &$E_s$ (kJ) \\
\midrule
\textbf{PS (Full-CSI)} &92.74  &101.28  &841.07  &126.87 &1161.96           &66.07\\
\textbf{PS (AASS-II)}  &91.51  &96.85   &830.72  &126.87 &1145.95           &65.20\\
\textbf{UAV (Full-CSI)} &90.40 &136.88  &769.51  &126.87 &1123.66           &64.41\\
\textbf{UAV (RAB)} &83.73  &115.12  &710.74  &126.87 &1036.46               &59.66\\
\textbf{Hybrid (Full-CSI)}  &83.04   &96.59  &723.72  &126.87 &1030.22      &59.16\\
\textbf{Hybrid (CSI-free)}  &79.23  &89.46   &683.93  &126.87 &\textbf{979.49} &\textbf{56.45}\\
\bottomrule
\end{tabular}
\end{table}

\section{Conclusion} \label{ConSec}
To enable large-scale, economical, and sustainable underground monitoring, we conceptualized a UAV-enabled WPUCN system, where a UAV is dispatched to charge a large number of remote UDs and collect sensor data. In this study, we modeled the system energy consumption and considered three WET approaches (i.e., traditional PS, UAV-enabled WET, and hybrid approaches) along with the full-CSI and CSI-free schemes for efficient WET operation. Based on these WET approaches, we proposed a time allocation strategy that minimizes the UAV's energy consumption while satisfying the throughput requirements of each UD and assuring the complete data offloading to the HAP. Through extensive modeling of a realistic farming scenario, the numerical results revealed that the proposed hybrid approach outperforms conventional WET approaches. Notably, its performance gain was significantly affected by factors such as the number of antennas, flying distance, number of UDs, burial depths, and VWC of the soil. Furthermore, under the derived optimal time allocation, our proposed hybrid approach with the CSI-free schemes achieved the lowest UAV energy consumption across all WET mechanisms, where the HAP employed the AASS-II scheme and the UAV operated the RAB scheme with the appropriate number of antennas. This demonstrated that efficient WET can be achieved without the need for CSI acquisition. Overall, our work confirmed the feasibility and effectiveness of the proposed UAV-enabled WPUCN system and provided valuable insights for future experimentation and practical deployments of this novel concept.

Although smart agriculture is regarded as the representative use case in this study, the proposed time allocation strategy for the UAV-enabled WPUCN system can be generalized for other underground applications, such as underground pipeline monitoring and post-disaster rescue, by appropriately adjusting the channel model and system parameters. Herein, potential impairment sources, including non-uniform signal attenuation across soil layers, multipath fading caused by gravel, and the impact of underground infrastructure, need to be considered in the system model to more accurately characterize the system performance of those applications. Furthermore, the formulated problem can be extended to jointly optimize the UAV's speed, flight distance, and hovering position to further enhance system performance, while accounting for practical constraints such as the UAV’s battery capacity, the UDs’ energy storage limitations, and the entire operation period. 

\bibliographystyle{IEEEtran} 
\bibliography{ref}

\end{document}